\documentstyle[preprint,aps,floats,epsfig]{revtex}
\tightenlines
\begin{document}
\draft

\preprint{\vbox{\hbox{JLAB-THY-99-037}}}

\title{ Origin of Relativistic Effects in the Reaction D(e,e'p)n
at GeV Energies}

\author{S. Jeschonnek$^{(1)}$ and J. W. Van Orden$^{(1,2)}$}

\address{\small \sl 
(1) Jefferson Lab, 12000 Jefferson Ave, Newport News, VA 23606 \\
(2) Department of Physics, Old Dominion University, Norfolk, VA 23529}

\date{\today}
\maketitle

\begin{abstract}
In a series of recent publications, a new approach to the
non-relativistic reduction of the electromagnetic current operator in
calculations of electro-nuclear reactions has been introduced.  In one
of these papers \cite{relcur}, the conjecture that at energies of a
few GeV, the bulk of the relativistic effects comes from the current
and not from the nuclear dynamics was made, based on the large
relativistic effects in the transverse-longitudinal response.  Here,
we explicitly compare a fully relativistic, manifestly covariant
calculation performed with the Gross equation, with a calculation that
uses a non-relativistic wave function and a fully relativistic current
operator. We find very good agreement up to missing momenta of $400$
MeV/c, thus confirming the previous conjecture.  We discuss slight
deviations in cross sections for higher missing momenta and their
possible origin, namely p-wave contributions and off-shell effects.
\end{abstract}
\pacs{24.10.Jv,~25.30.Fj,~25.10+s}

\section{Introduction}
Currently, there is a lot of activity in the field of electron
scattering from nuclei at a few GeV. There exists a broad experimental
program of electron scattering studies at Jefferson Lab, MAMI, and
Bates, aimed at understanding the short range structure of nuclei and
the properties of nucleons in the nuclear medium.  It would be ideal
to perform theoretical calculations of these processes in a completely
microscopic fashion, starting from a Lagrangian. Such an approach
would contain explicit relativistic treatments of the nuclear dynamics
and the electromagnetic current, as well as a consistent treatment of
the initial nuclear state and the in general quite complicated
hadronic final state. In practice, for high-energy
inelastic electron scattering, it is very difficult to calculate both
the nuclear ground state and the final hadronic scattering state
consistently and it is unlikely that a consistent, fully microscopic
and, therefore, relativistic treatment will be available for medium and
heavier nuclei in the near future.  
Currently, there are several
approximate calculations for particular electro-nuclear reactions
available for few-body systems \cite{vano,tjon}. However, it is
difficult to extend these approaches for the few-body systems to
energies above the pion emission threshold. With the considerable
effort one expects will be put into this field, in the next few years
calculations for the few GeV regime for reactions with deuteron
targets will hopefully become available.  For heavy nuclei, there are
only relativistic mean-field calculations available, which do not allow
for the investigation of ground-state correlations.

Therefore, it is an essential question if the main physical features
of high energy electro-nuclear reactions can be incorporated in
theoretical calculations in an effective manner, which would allow one
to perform these calculations for a variety of target nuclei in the
few GeV regime. One essential feature at GeV energies is relativity,
and this is the topic with which we concern ourselves in this paper.

Relativistic effects show up in three places: in the kinematics, in
the electromagnetic current, and in the nuclear/hadronic dynamics.
The first point, relativistic kinematics, is trivial. In order to
treat the nuclear dynamics in a relativistic way, one can solve
the Bethe Salpeter equation or one of its variations, e.g. the Gross
equation \cite{grosseq}. The electromagnetic current has traditionally
been used in a non-relativistic reduction, which assumes that the
transferred momentum, transferred energy and initial nucleon momentum
are small compared to the nucleon mass. These assumptions are not
valid for modern day experiments with GeV electron beams and energy
and momentum transfers of 1 GeV or more, and there exist several
improved schemes.  In a series of recent publications
\cite{relcur,pvquique,quiqueinc,quiquecoin} a new approach to the
problem of the non-relativistic reduction of the current was
introduced, and in \cite{relcur}, it was conjectured that the bulk of
the relativistic effects at energies of a few GeV stem from the
current, not from the nuclear dynamics. This conjecture is based on
the large enhancement of the transverse-longitudinal response observed
in this scheme, which was also seen in the fully relativistic
calculation of \cite{tjon}. If this conjecture is indeed true,
relativity could be taken into account by applying the operator
introduced in \cite{relcur,pvquique,quiqueinc,quiquecoin} together
with non-relativistic wave functions to the calculation of
electro-nuclear reactions. This would obviously be very useful.

Although strong evidence was presented in \cite{relcur}, this
conjecture can only be confirmed by a direct comparison to a fully
relativistic, microscopic calculation. This comparison is carried out
in the following sections for the deuteron target. As we wish to
concentrate for the moment on the problem of relativity only, we
perform this comparison for Plane Wave Impulse Approximation (PWIA).
 A realistic calculation must treat final
state interactions, too, but the conjecture can be tested successfully
at the plane wave level. Moreover, a fully microscopic, consistent
calculation of final state interaction at GeV energies is extremely
difficult and not available at the moment. The high energy final state
interactions are usually treated within the framework of Glauber
theory \cite{sofsi}.

\section{Brief review of formalism and notation}

We start by introducing some notation and giving a brief summary
of the basic formalism of $(e,e'p)$ reactions. More details can be
found in \cite{raskintwd,dmtrgross}.

The differential cross section in the lab frame is 

\begin{eqnarray}
\left ( \frac{ d \sigma^5}{d \epsilon' d \Omega_e d \Omega_N}
\right ) _{fi}^h  & = & 
\frac{m_N \, m_f \, p_N}{8 \pi^3 \, m_i} \, \sigma_{Mott} \, 
f_{rec}^{-1} \, \nonumber \\
& & \Big[ \left ( v_L R_{fi}^L +   v_T R^T_{fi}
 + v_{TT} R_{fi}^{TT} + v_{TL} R_{fi}^{TL} \right )
  \nonumber \\
& & +  h \left ( v_{T'} R_{fi}^{T'} +  v_{TL'} R_{fi}^{TL'}
\right ) \Big] \, ,
\label{wqdef}
\end{eqnarray}
where $m_i$, $m_N$ and $m_f$ are the masses of the target nucleus, the
ejectile nucleon and the residual system, $p_N$ and $\Omega_N$ are the
momentum and solid angle of the ejectile, $\epsilon'$ is the energy of
the detected electron and $\Omega_e$ is its solid angle.  The helicity
of the electron is denoted by $h$.  
The Mott cross section is 
\begin{equation}
\sigma_{Mott} = \left ( \frac{ \alpha \cos(\theta_e/2)}
{2 \varepsilon \sin ^2(\theta_e/2)} \right )^2
\end{equation}
and the recoil factor is given by
\begin{equation}
f_{rec} = | 1+ \frac{\omega p_x - E_x q \cos \theta_x}
{m_i \, p_x} | \, .
\end{equation}
The coefficients $v_K$ are the
leptonic coefficients, and the $R_K$ are the response functions
which are defined by

\begin{eqnarray}
R_{fi}^L & \equiv & | \rho (\vec q)_{fi}|^2 \nonumber \\
R_{fi}^T & \equiv & | J_+ (\vec q)_{fi}|^2 
+ | J_- (\vec q)_{fi}|^2  \nonumber  \\
R_{fi}^{TT} & \equiv &  2 \, \Re \, \big[ J_+^* (\vec q)_{fi} \,
J_- (\vec q)_{fi} \big] \nonumber  \\
R_{fi}^{TL} & \equiv & - 2 \, \Re \, \big[ \rho^* (\vec q)_{fi} \,
( J_+ (\vec q)_{fi} - J_- (\vec q)_{fi}) \big] \nonumber \\
R_{fi}^{T'} & \equiv & | J_+ (\vec q)_{fi}|^2 -
 | J_- (\vec q)_{fi}|^2  \nonumber  \\
R_{fi}^{TL'} & \equiv & - 2 \, \Re \, \big[ \rho^* (\vec q)_{fi} \, 
( J_+ (\vec q)_{fi} + J_- (\vec q)_{fi}) \big] \, , 
\label{defresp}
\end{eqnarray}
where the $J_{\pm}$ are the spherical components of the current.  For
our calculations, we have chosen the following kinematic conditions:
the z-axis is parallel to $\vec q$, the missing momentum is defined as
$\vec p_m \equiv \vec q - \vec p_N$, so that in Plane Wave Impulse
Approximation (PWIA), the missing momentum is equal to the negative
initial momentum of the struck nucleon in the nucleus, $\vec p_m =
-\vec p$. We denote the angle between $\vec p_m$ and $\vec q$ by
$\theta$, and the term ``parallel kinematics'' indicates $\theta =
0^o$, ``perpendicular kinematics'' indicates $\theta = 90^o$, and
``anti-parallel kinematics'' indicates $\theta = 180^o$.  Note that
both this definition of the missing momentum and the definition with
the other sign are used in the literature.  In this paper, we assume
that the experimental conditions are such that either the kinetic
energy of the outgoing nucleon and the angles of the missing momentum,
$\theta$, and the azimuthal angle $\phi$, are fixed, or that the
transferred energy $\omega$, the transferred momentum $\vec q$, and
the azimuthal angle $\phi$, are fixed. In the former case, the
transferred energy and momentum change for changing missing momentum,
in the latter situation, the kinetic energy and polar angle of the
outgoing proton change for changing missing momentum.

The electromagnetic current operator
\begin{equation}
J^\mu (P\Lambda ;P^{\prime }\Lambda ^{\prime })=\bar u(P^{\prime }\Lambda
^{\prime })\left[ F_1\gamma ^\mu +\frac i{2m_N}F_2\sigma ^{\mu \nu }Q_\nu
\right] u(P\Lambda )  \label{defcur} \,
\end{equation}
where $P, P'$ indicate the four-momenta of the nucleon,
can be rewritten in a form that is more suitable for application to nuclear
problems:
\begin{equation}
J^\mu (P\Lambda ;P'\Lambda ^{\prime }) \equiv 
\chi _{\Lambda
^{\prime }}^{\dagger }  \, \, \bar{J}^{\mu }(P;P^{\prime }) 
\, \, \chi _{\Lambda }^{{}}
\end{equation}
with
\begin{eqnarray}
\bar J^o &=& \rho = f_o \left (
\xi _o + \, i \, \, \xi _o^{\prime } \, \left( \vec q \times \vec p
\right) \cdot \vec{\sigma }   \right ) \nonumber \\
\bar J^3 &=& \, \,  \frac{\omega}{q} \, \, \bar J^o  \nonumber \\
\bar J^{\bot } &=& f_o \left (  \xi _{1} \left[ \, \vec p
-\left( \frac{\vec q
\cdot \vec p}{q ^2}\right) \vec q \, \right] \right. \\
&-i&
\left\{ \xi _1^{\prime } \left( \vec q \times \vec{\sigma }
\right)
% \right.  \right. \nonumber\\
  +  \left.  \xi _2^{\prime } \left( \vec q \cdot \vec{\sigma }
\right) \left( \vec q \times \vec p \right) +\xi
_3^{\prime }\left[ \left( \vec q \times \vec p \right)
\cdot \vec{\sigma }\right] \left[ \vec p -\left( \frac{\vec q 
\cdot \vec p }{q ^2}\right) \vec q \right]
\right\} \right )  .
\end{eqnarray}
Here, $f_o, \xi_i, \xi_i'$ are all functions of $\omega, q, p^2$;
their explicit forms are given in \cite{relcur}. For the reasons
explained in \cite{relcur}, we refer to the operator associated with
$\xi_o$ as zeroth-order charge operator, we call the term containing
the $\xi_o'$ first-order spin-orbit operator, the term containing
$\xi_1$ first-order convection current, the term containing $\xi_1'$
zeroth-order magnetization current, the term containing $\xi_2'$
first-order convective spin-orbit term, and the term containing
$\xi_3'$ second-order convective spin-orbit term. 

Note that there are more terms in the full current than with the
commonly used strict non-relativistic reduction, which assumes that the
transferred momentum $q$ is smaller than the nucleon mass, and that
both the initial nucleon momentum and the transferred energy are
smaller than $q$ and therefore much smaller than the nucleon mass.
Under these assumptions, the current operator simplifies to the form
\begin{eqnarray}
\bar J^o_{nonrel} &=& G_E \nonumber \\
\bar J^{\perp}_{nonrel} &=& - \frac{i}{2 \, m_N} \,  G_M \, \left (
\vec q \times \vec \sigma \right ) + \frac{1}{m_N} \, G_E \,
\left ( \vec p - \frac{\vec q \cdot \vec p}{q^2} \, \vec q
\right ) \nonumber \,,
\end{eqnarray}
which contains only the zeroth-order charge operator, the zeroth-order
magnetization current and the first-order convection current.

In non-relativistic Plane Wave Impulse Approximation (PWIA), we have
\begin{equation}
\frac{ d^5 \sigma}{d \epsilon' d \Omega_e d \Omega_N} = 
\frac{m_N \, m_f \, p_N}{m_i} \, \sigma_{eN} \, 
f_{rec}^{-1} \, n (\vec p) \,,
\label{pwiafactor}
\end{equation}
where $n(\vec p)$ is the momentum distribution, and the
$eN$ cross section is given by
\begin{equation}
\sigma_{eN} = \sigma_{Mott} \sum_k v_k R_k^{single \, nucleon}
\end{equation}
and the single nucleon responses are related to the nuclear responses
by
\begin{equation}
R_K^{nucleus} = (2 \pi )^3 \, R_k^{single \, nucleon} \, n(\vec p)
\end{equation}
so that one has in total:
\begin{equation}
\frac{ d^5 \sigma}{d \epsilon' d \Omega_e d \Omega_N} = 
\frac{m_N \, m_f \, p_N}{m_i} \,
f_{rec}^{-1} \,  \sigma_{Mott} \, n (\vec p) \,
\sum_k v_k R_k^{single \, nucleon}
\end{equation}
The momentum distribution is simply the Fourier transform of
the wave function, and in the non-relativistic case takes the form:
\begin{equation}
n(\vec p) = \frac{1}{2 \pi^2} ( u(p)^2 + w(p)^2)
\label{momdis}
\end{equation}
where $u(p), w(p)$ are the S-wave and D-wave
 functions in momentum space, and the normalization condition is
\begin{equation}
\int d^3 \vec p \, n(\vec p) = 1
\label{nrnorm}
\end{equation}
Note that the different parts of the wave function do not interfere as
long as the target deuteron is unpolarized.  The above formulas, and
especially the factorization into $eN$ cross section and momentum
distribution, is valid only in non-relativistic PWIA, not in PWBA. The
Born approximation for $(e,e'p)$ scattering contains in addition to
the graph where the photon couples to to the proton the graph where
the photon couples to the neutron.  The difference between PWIA and
PWBA for high energy transfers is small, as the second contribution
involves the deuteron wave function taken at the momentum of the
outgoing proton, which is high, and therefore the value of the
deuteron wave function is very small. However, the PWBA is sufficient
to break factorization and to permit the interference of the different
wave function components. In the following, we present explicit
analytic formulas for PWIA; these formulas are considerably simpler
than the corresponding PWBA expressions. For consistency, the
numerical results are also presented in PWIA. Their deviation from the
PWBA results at the kinematics used here is very small, on the order
of a few percent. None of our results and conclusions are affected by
this.

For the full current operator, the single-nucleon
responses take the following form:

\begin{eqnarray}
R_L^{sn} &=& f_o^2 (\xi_o^2 + \kappa^2 \delta^2 \xi_o'^2)
\nonumber \\ 
&=& \frac{\kappa ^2}{\tau} ( G_E ^2 + \delta ^2 W_2) 
\nonumber \\ 
R_T^{sn} &=& f_o^2 (2 \kappa^2 \xi_1'^2 + \kappa^4 \delta^2 \xi_2'^2
+ \delta^2 \xi_1^2 + \kappa^2 \delta^4 \xi_3'^2 - 2 \kappa^2 
 \delta^2 \xi_1' \xi_3') \nonumber \\ 
&=& 2 W_1 + \delta ^2 W_2 \nonumber \\ 
R_{TT}^{sn} &=&  f_o^2 (\kappa^4 \delta^2 \xi_2'^2 + 2 \kappa^2 
 \delta^2 \xi_1' \xi_3' -  \delta^2 \xi_1^2 
- \kappa^2 \delta^4 \xi_3'^2) \cos (2 \varphi) \nonumber \\ 
&=& -\delta ^2 W_2 \cos (2 \varphi) \nonumber \\ 
R_{TL}^{sn} 
&=& 2 \sqrt{2} \cos (\varphi) f_o^2 (\delta \xi_o \xi_1
+\kappa^2 \delta \xi_o' (\xi_1' - \delta^2 \xi_3'))
\nonumber \\ 
&=& 2 \sqrt{2} \cos (\varphi)\frac{\kappa}{\sqrt{\tau}} 
\sqrt{ 1 + \tau +\delta ^2} \delta  W_2 
\label{snrex}
\end{eqnarray}

with the abbreviations $W_1 \equiv \tau G_M^2$ and
$W_2 \equiv \frac{1}{1+\tau} ( G_E^2 + \tau G_M^2)$.

For the strict non-relativistic reduction, the single nucleon
responses are given by:
\begin{eqnarray}
R_L^{sn} &=& G_E ^2
\nonumber \\ 
R_T^{sn} &=& 2 \kappa^2 \, G_M^2  + \delta ^2 G_E^2 \nonumber \\ 
R_{TT}^{sn} &=& -\delta ^2 G_E^2 \cos (2 \varphi) \nonumber \\ 
R_{TL}^{sn} &=& 
\delta  G_E^2 2 \sqrt{2} \cos (\varphi)
\label{snrnr}
\end{eqnarray}

The dimensionless variables are defined as follows:
\begin{eqnarray}
\kappa &=& \frac{|\vec q|}{ 2 m_N} \nonumber \\ 
\delta &=& \frac{p_{\perp}}{m_N} \nonumber \\ 
\tau &=& \kappa ^2 - \lambda ^2 \nonumber \\ 
\lambda &=& \frac{\omega}{2 m_N}
\end{eqnarray}

\begin{figure}[!h,t]
\begin{center}
\leavevmode
\epsfig{file = 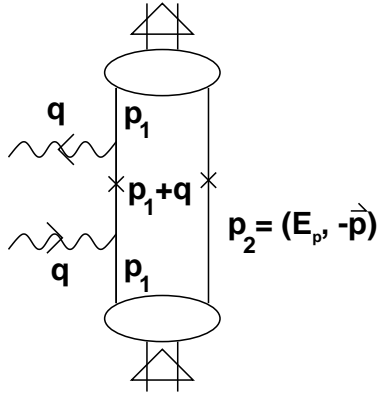, height=9cm}
\end{center}
\vspace{-2cm}
\caption{This figure shows the relevant kinematic quantities for the
electrodisintegration of the deuteron in PWIA.}
\label{graph}
\end{figure}

The above formulas for the non-relativistic PWIA do not hold for a
relativistic treatment of the situation. Besides the presence of
singlet and triplet P waves, which appear only in a full relativistic
treatment of the hadronic state, the factorization breaks down.  This
fact was observed earlier, e.g. in \cite{wallynf,rpwia,gardner}, and can be
attributed to the presence of negative energy states in the
relativistic treatment. To illustrate this point, we write out the
expression for the hadronic tensor $W^{\mu \nu}$ in the $D(e,e'p)n$
reaction in the relativistic plane wave impulse approximation
(RPWIA). The relevant kinematic quantities are identified in
Fig. \ref{graph}; in the following, $\check p$ indicates a four-vector
with on-shell energy momentum relation, i.e. $p^o = E_p := \sqrt{\vec
p \, ^2 + m^2}$:
\begin{eqnarray}
W^{\mu \nu} &=& \frac{1}{3}  \mbox{Tr} ( J^{\mu} (-q) S_F (p_1 + q)
J^{\nu} (q) \nonumber \\
& & \sum_{\lambda_d \lambda_2} \sum_{\lambda_1 ' \lambda_1}
[ u(\vec p_1, \lambda_1) \Psi^+_{\lambda_1 \lambda_2 \lambda_d}
(\vec p_1) + v (-\vec p_1, \lambda_1)  \Psi^-_{\lambda_2
\lambda_1 \lambda_d} (\vec p_1)]
\nonumber \\
& & [ \bar u(\vec p_1, \lambda_1') 
\Psi^{+ *}_{\lambda_2 \lambda_1' \lambda_d} (\vec p_1) + 
\bar v (-\vec p_1, \lambda_1') \Psi^{- *}_{\lambda_2
\lambda_1' \lambda_d} (\vec p_1)] ) \,
\label{hadtens}
\end{eqnarray}
where the wave function components are given by
\begin{eqnarray}
 \Psi^+_{\lambda_2 \lambda_1 \lambda_d} (\vec p) &=&
\frac{1}{\sqrt{8 \pi}} D_{\lambda_d \lambda_1 \lambda_2}^{1 *}
(\varphi,\vartheta, -\varphi) [ (u(p) + \sqrt{2} w(p))
\delta_{\lambda_1, \lambda_2} + (\sqrt{2} u(p) - w(p))
\delta_{\lambda_1, -\lambda_2}] \nonumber \\
 \Psi^-_{\lambda_2 \lambda_1 \lambda_d} (\vec p) &=& -
\frac{1}{\sqrt{8 \pi}} D_{\lambda_d \lambda_1 -\lambda_2}^{1 *}
(\varphi,\vartheta, -\varphi) 2 \lambda_2 (\sqrt{3} v_s(p)
\delta_{\lambda_1, \lambda_2} + \sqrt{3} v_t (p) 
\delta_{\lambda_1, -\lambda_2}) \,,
\end{eqnarray}
where $D^1$ denotes the $J=1$ Wigner D function, and the full wave
function is
\begin{equation}
\Psi _{\lambda_2, \lambda_d} (\vec p) = \sum_{\lambda_1} 
[ u(\vec p_1, \lambda_1)  \Psi^+_{\lambda_1 \lambda_2 \lambda_d}
(\vec p_1) + v (-\vec p_1, \lambda_1) \Psi^-_{\lambda_2
\lambda_1 \lambda_d} (\vec p_1)] \,.
\end{equation}
Note that whenever the P wave contributions $v_s$ and $v_t$ present in
$\Psi^-$ appear in the hadronic tensor eq. (\ref{hadtens}), they are
multiplied by a negative energy spinor $v$.

After some algebra, the hadronic tensor can be rewritten as
\begin{equation}
W^{\mu \nu} = \frac{2 M_d}{3} \mbox{Tr} ( J^{\mu} (-q) 
\Lambda^+ (p_1 + q) J^{\nu} (q) N) \,,
\end{equation}
where $\Lambda^+$ is the positive energy projection operator and $N$
is the sum of a time-like vector, a space-like vector, and a scalar
component:
\begin{equation}
N = \frac{3}{ 16 \pi} [ n_{TV} (p) \frac {P \cdot \check{p}_1}
{m M_d^2} \not P + n_{SV} (p) \frac{\not k}{m} + n_S (p) ] \,
\end{equation}
with the three distributions given by:
\begin{eqnarray}
n_{TV} (p) &=& u^2 (p) + w^2 (p) + v_s^2 (p) + v_t^2 (p) \\
n_{SV} (p) &=&  u^2 (p) + w^2 (p) -  v_s^2 (p) - v_t^2 (p)
\nonumber \\ 
&-& \frac{2 m}{\sqrt{3} p} [(u(p) + \sqrt(2) w(p)) v_s (p)
- ( \sqrt(2) u(p) - w(p)) v_t(p)] \\
n_S (p) &=& u^2 (p) + w^2 (p) - v_s^2 (p) - v_t^2 (p) \nonumber \\
&+& \frac{2 p}{\sqrt{3} m} [(u(p) + \sqrt(2) w(p)) v_s (p)
- ( \sqrt(2) u(p) - w(p)) v_t(p)] \, .
\label{defmds}
\end{eqnarray}
The four vectors used here are the deuteron four-momentum $P$ and $k =
(0,\vec p)$, where $\vec p$ is the three momentum component of
$p_1$. From these formulas it is obvious that apart from the time-like
vector distribution, there is indeed interference between the P waves
and the S and D waves in RPWIA. As already noted above, this
interference stems from the presence of the negative energy terms in
eq. (\ref{hadtens}), and is therefore a genuinely relativistic effect.
The normalization condition,
\begin{equation}
2 M_d \delta_{\lambda_d, \lambda_d'} = \int \frac{d^3 p}{(2 \pi)^3} 
\frac{m}{E_p} \sum_{\lambda_2} \bar \Psi_{\lambda_2, \lambda_d'}
( \vec p) \gamma^o \Psi_{\lambda_2, \lambda_d} (\vec p)
\label{wallynormcond}
\end{equation}
leads to the condition:
\begin{equation}
1 = \frac{1}{3}  \int \frac{d^3 p}{(2 \pi)^3} \frac{m}{E_p}
\mbox{Tr} (\gamma^o N) \,.
\end{equation}
Note that the normalization condition in eq. (\ref{wallynormcond}) is
a relativistic normalization condition and differs from the
non-relativistic normalization condition in eq. (\ref{nrnorm}).  These
differences do not matter as long as consistent phase space factors
are taken into account in the calculation of the cross section.  It is
clear that only the time-like vector momentum distribution term will
give a non-vanishing contribution to the trace in the integrand, so
that we obtain the familiar normalization condition:
\begin{equation}
1 = \int \frac{dp p^2}{2 \pi^3} \left (
u^2 (p) + w^2 (p) + v_s^2 (p) + v_t^2 (p) \right )
\end{equation}

\begin{figure}
\begin{center}
\vspace{-2cm}
\leavevmode
\epsfig{file = 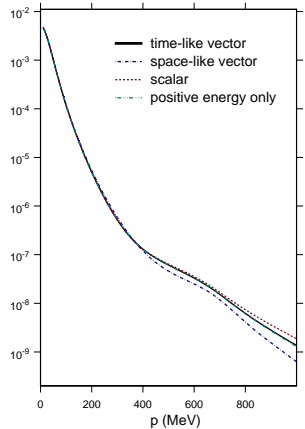, height=12cm}
\end{center}
\vspace{-5cm}
\caption{We show the different momentum distributions as
defined in eq. (\protect \ref{defmds}).  The following
distributions are presented: the time-like vector distribution (solid
line), the space-like vector distribution (dash-dotted line), the
scalar distribution (dashed line), and the time-like vector
distribution with the P-waves put to zero (dash-double-dotted line).}
\label{figmdsa}
\end{figure}

\begin{figure}
\begin{center}
\vspace{-2cm}
\leavevmode
\epsfig{file = 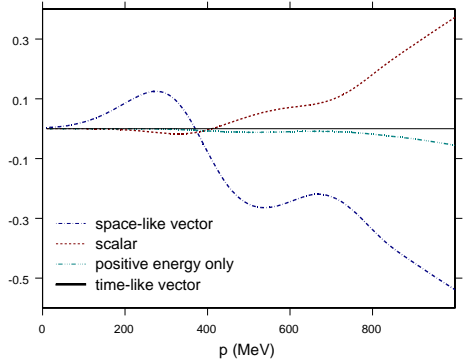, height=12cm}
\end{center}
\vspace{-5cm}
\caption{We plot the
ratio $R = \frac{n_x - n_{TV}}{n_{TV}}$. The curves indicate the same
distributions as in Fig. \ref{figmdsa}.}
\label{figmdsb}
\end{figure}

From the hadronic tensor, one can compute the response functions by
picking the appropriate values of the indices $\mu, \nu$ and
evaluating the traces.  The explicit expressions for the responses are
given in appendix \ref{apa}.  In this process, not only the time-like
vector momentum distribution, but also the scalar and the space-like
momentum distributions enter in the evaluation of the response
functions. In order to give an idea about the amount of interference
that is possible in the response functions, we have plotted the three
different momentum distributions in
Figs. \ref{figmdsa} and \ref{figmdsb}. In Fig. \ref{figmdsa}, we show the
time-like vector distribution (solid line), the space-like vector
distribution (dash-dotted line), and the scalar distribution (dashed
line). In addition, we have plotted the time-like vector distribution
with the P-waves put to zero (dash-double-dotted line).  This would
correspond to the non-relativistic momentum distribution.  In
Fig. \ref{figmdsb}, we show the ratio R, $R = \frac{n_x -
n_{TV}}{n_{TV}}$, of the deviations of the different distributions
from the time-like distribution to the time-like vector
distribution. One can see clearly that both the space-like vector and
scalar distributions deviate significantly from the time-like vector
distribution, in the case of the space-like vector distribution, this
happens already for rather small momenta.  For increasing momenta, the
interference effects grow considerably, until they reach 30 \% to
50\%. In contrast, the effect of the P waves in the time-like vector
distribution is small, and increases only slightly for very high
momenta, $p > 800$ MeV/c. This is as expected, as the P waves do not
interfere with S or D waves in this case, and their normalization is
too small to make them significant by themselves, compared to the S
wave or D wave contribution.

\section{Comparison of different approaches to relativity}

In this section, we compare the results of a plane wave calculation of
the $D(e,e'p)n$ reaction in the framework of the Gross equation
\cite{grosseq,vano} (Model R) with a calculation using a
non-relativistic nucleon-nucleon potential, namely the Argonne V18
potential \cite{av18}, and the full relativistic current operator
(Model A1). For an overview of all employed models and naming
conventions, see appendix \ref{apnames}. Both models give a very good
fit to the $N N$ scattering data. At the high energy and momentum
transfer we consider here, the Plane Wave Impulse Approximation and
the Plane Wave Born Approximation become almost identical, they differ
by 2\% or 3\% at most.  In order to be consistent with the analytic
formulas presented in the previous section and in the appendix, the
calculations we show are done in PWIA. We perform our comparison for
the basic observable, the differential cross section.

\begin{figure}[!h,p]
\begin{center}
\leavevmode
\epsfig{file = 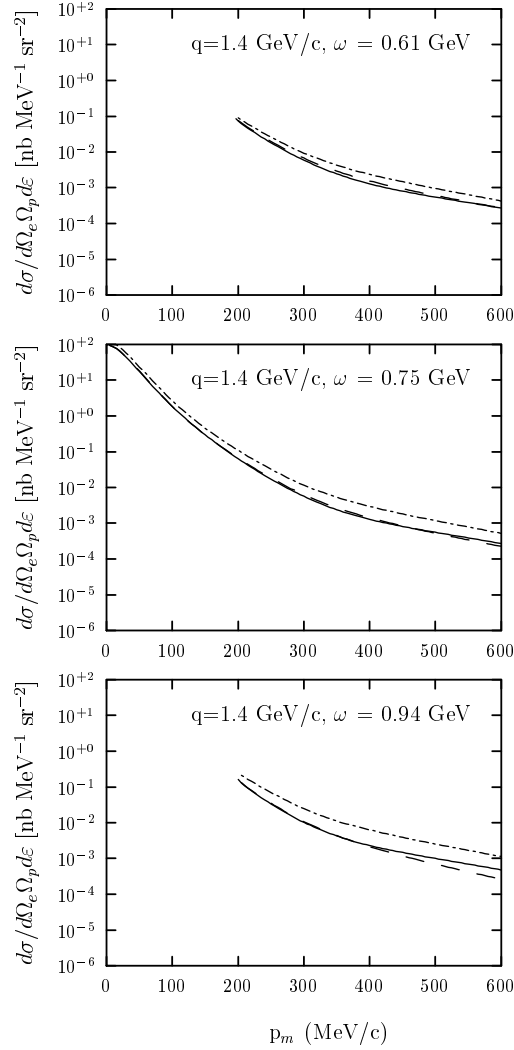, height = 20cm}
\end{center}
\vspace{-3cm}
\caption{The differential cross section for the $^2H(e,e'p)n$ reaction
is shown for fixed 3-momentum transfer $q = 1.4$ GeV/c and different
values of the fixed energy transfers $\omega$: 0.61 GeV (a), 0.75 GeV
(b), and 0.94 GeV (c).  The kinematic conditions in the middle panels
correspond to quasi-free conditions. The solid line shows the result
of the fully relativistic calculation (Model R), the dashed line shows
the result of the calculation with the Argonne V18 wave function and
the full relativistic current operator (Model A1), and the dash-dotted
line shows the result of the calculation with the Argonne V18 wave
function and the strict non-relativistic reduction of the current
operator (Model NR).}
\label{fig1}
\end{figure}

\begin{figure}[!h,t,p]
\begin{center}
\vspace{-3cm}
\leavevmode
\epsfig{file = 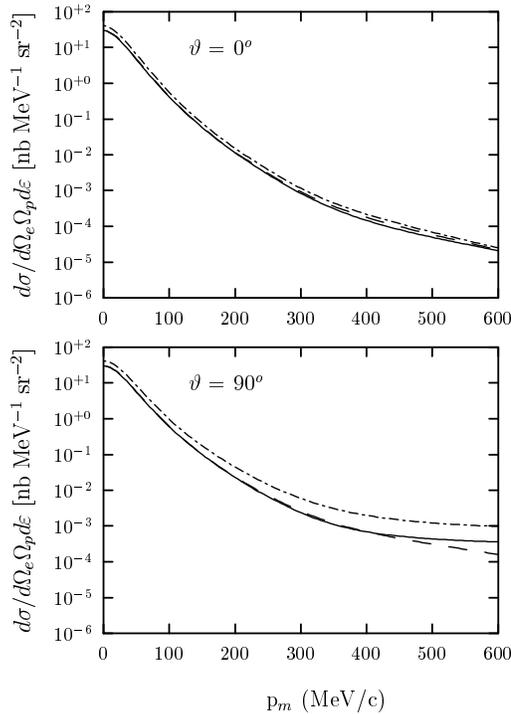, height=20cm}
\end{center}
\vspace{-8cm}
\caption{The differential cross section for the $^2H(e,e'p)n$ reaction
is shown for fixed kinetic energy, $T_{kin} = 1$ GeV, of the outgoing
proton in parallel and perpendicular kinematics. The curves have the same 
meaning as in Fig. \ref{fig1}.}
\label{fig1a}
\end{figure}

In Figs. \ref{fig1} and \ref{fig1a}, we compare the fully relativistic,
covariant results (Model R) with the results for the Argonne V18 wave
function in conjunction with the full relativistic current operator
(Model A1).  We have also added a curve showing the results obtained
with the Argonne V18 wave function and the traditional, strictly
non-relativistic current operator (Model NR), in order to give an idea
of the actual size of relativistic effects in the current. In the
first figure, we have fixed the transferred momentum, $q$, to $q =
1.4$ GeV/c, and the transferred energy, $\omega$, to a value
corresponding to the quasi-elastic peak, $\omega = 0.75$ GeV (middle
panel), to a value below the quasi-elastic peak, $\omega = 0.61$ GeV
(top panel), and to a value above the quasi-elastic peak, $\omega =
0.94$ GeV (bottom panel).  This choice corresponds to values of the
$y$-scaling variable of $y = 0$ GeV/c, $y= -0.2$ GeV/c, and $y=0.2$
GeV/c. The $y$ variable is defined as the negative minimal missing
momentum, for details, see e.g.  \cite{yscal}.  The solid line shows
the fully relativistic calculation, the dashed line shows the
calculation with the relativistic current operator and the AV18 wave
function, and the dash-dotted line shows the the calculation with the
traditional non-relativistic reduction of the current and the AV18
wave function.  We start our discussion with the lowest energy
transfer, $\omega = 0.61$ GeV. One sees that the fully relativistic
calculation and the AV18 with full relativistic current are relatively
close to each other, with the fully relativistic result a bit lower
for missing momenta around $400$ MeV/c.  The result obtained with the
non-relativistic current operator is larger than the relativistic
curves.  For the quasi-free kinematics (middle panel), the fully
relativistic calculation and the AV18 plus full relativistic current
agree quite nicely, with the fully relativistic curve being slightly
lower for momenta around $350$ MeV/c and then being slightly higher
for momenta higher than $500$ MeV/c. The non-relativistic result is
significantly higher for all missing momenta. This trend continues for
the kinematics above the quasi-elastic peak (lower panel), where the
difference between non-relativistic and relativistic curves increases
again. This is to be expected, as one of the basic assumptions of the
strictly non-relativistic reduction is that the transferred energy is
much smaller than the transferred momentum. This condition is
fulfilled to a certain extent below the quasi-elastic peak, but not on
or above it.  The fully relativistic result and the AV18 plus full
relativistic current agree for missing momenta up to $400$ MeV/c, then
the AV18 result drops off a bit.  This difference is the largest
between these two calculations which we have encountered so far, and
it is logical that it should arise where it does: certainly,
relativistic effects are expected to be strongest for situations where
both energy and momentum transfer are large, and for high missing
momenta, which correspond in the absence of final state interactions
to the initial momentum of the nucleon in the nucleus. It is logical
to expect any relativistic effect in the nuclear dynamics to show up
at high missing momenta.  Below, we will discuss the relativistic
effects surpassing kinematics and current.

In order to get a different view, we also show the same results for a
different kinematic situation: now, we fix the kinetic energy of the
outgoing proton to 1 GeV, and consider parallel and perpendicular
kinematics. The results are shown in Fig. \ref{fig1a}.  For parallel
kinematics (top panel), the fully relativistic result and the AV18
plus full relativistic current result coincide for missing momenta below $300$
MeV/c, for higher missing momenta, the fully relativistic curve is a
bit lower. The non-relativistic result differs from the other results
already at the lowest missing momenta, but the difference is
comparatively small in these kinematics, the momentum transfer
increases very quickly, so that the relation $\omega << q$ is roughly
applicable, as in the case of Fig. \ref{fig1}, top panel.  In
perpendicular kinematics, the fully relativistic result and the AV18
plus exact current result agree very nicely up to $p_m = 450$ MeV/c,
for higher missing momenta, the two calculations start to diverge,
leading to a difference similar to the one observed for the highest
energy transfer in the fixed q kinematics.

To summarize the results of our comparison so far, we have seen that
there is very good overall agreement between the fully relativistic
result (Model R) and the AV18 plus full relativistic current result
(Model A1), with some slight deviations at higher missing momenta,
$p_m \geq 450$ MeV/c, especially for energy transfers $\omega$
comparable to q. The calculation with the AV18 wave function and the
non-relativistic reduction of the current (Model NR) differs from the
relativistic results considerably, starting from the lowest missing
momenta. In the light of these results, it is fair to say that the
bulk of the relativistic effects in the few GeV region does stem from
the current, as conjectured in \cite{relcur}.

\begin{figure}
\begin{center}
\vspace{-4cm}
\leavevmode
\epsfig{file = 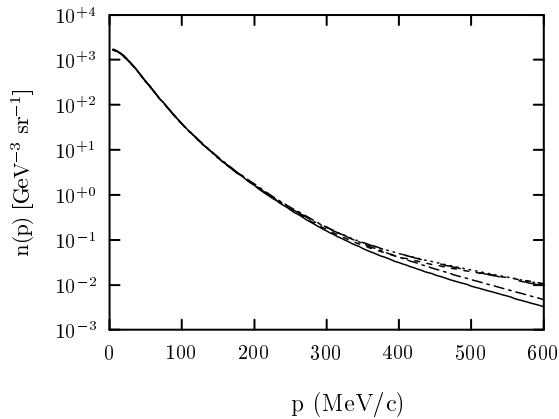, height=22cm}
\vspace{-14cm}
\end{center}
\caption{The momentum distribution n(p) for the deuteron is shown for
several wave function models. The solid line shows the Bonn result,
the dashed line shows the Paris result, the dash-dotted line shows the
CD Bonn result, and the long dashed line shows the Argonne V18 result.
The short-dashed line shows the momentum distribution calculated with
the S wave and D wave of the Gross wave function.}
\label{figmomdis}
\end{figure}

In order to ascertain the influence of relativity in contrast to the
influence of the modeling of the $N N$ interaction properly, we carry
out an additional comparison. We take a parameterization of the S wave
and D wave part of the Gross wave function, and use this together with
the full relativistic current (Model G1).  Whether the model for the
$N N$ interaction is relativistic or not, this choice alone does not
specify the model. The data constrain the deuteron wave function
somewhat, but it is well known that for higher momenta, where the D
wave is dominant, different models make different
predictions. E.g. the Bonn \cite{bonn} and the Paris \cite{paris} wave
functions are both non-relativistic, but differ greatly in their D
wave content. In our comparison, we are interested in the different
treatment of relativity, and want to eliminate differences due to the
difference in the $N N$ interaction modeling. We therefore show a plot
of the momentum distribution as given by eq. (\ref{momdis}) of several
non-relativistic models, and also plot the momentum distribution
obtained with the S wave and D wave of the Gross wave function.  We
omit the P waves here in order to compare the same quantity, the
different normalization induced by this has been taken into account.
As seen in Fig. \ref{figmomdis}, all wave function models agree for
momenta up to 300 MeV/c, and then start to diverge. The two Bonn
potential curves, Bonn \cite{bonn} and CD Bonn \cite{machleidt}, are
lower than Paris, Argonne V18 and the Gross wave function.  The latter
three are fairly close, with the Paris and Argonne V18 slightly higher
than the Gross wave function.  This means that our comparison between
the fully relativistic calculation (Model R) and the Argonne V18 with
the full relativistic current (Model A1) is meaningful, as the
differences in these calculations should arise mainly from the
different treatment of relativity, not from the differences in the
wave function. Note that a comparison of the fully relativistic
calculation with one of the Bonn potential wave functions plus full
relativistic current would not have achieved that.

\begin{figure}
\begin{center}
\leavevmode
\epsfig{file = 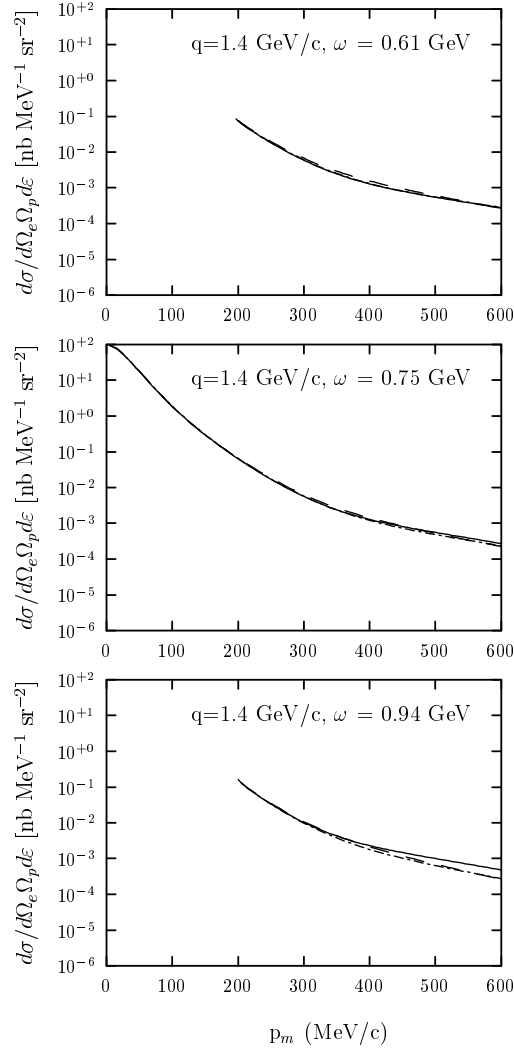, height=20cm}
\end{center}
\vspace{-3cm}
\caption{The differential cross section for the $^2H(e,e'p)n$ reaction
is shown for fixed 3-momentum transfer $q = 1.4$ GeV/c and different
values of the fixed energy transfers $\omega$: 0.61 GeV (a), 0.75 GeV
(b), and 0.94 GeV (c).  The kinematic conditions in the middle panels
correspond to quasi-free conditions. The solid line shows the result
of the fully relativistic calculation (Model R), the dashed line shows
the result of the calculation with the Argonne V18 wave function and
the full relativistic current operator (Model A1), and the dash-dotted
line shows the result of the calculation with the Gross S wave and D
wave plus the full relativistic current operator (Model G1).}
\label{fig2}
\end{figure}

\begin{figure}
\begin{center}
\vspace{-3cm}
\leavevmode
\epsfig{file = 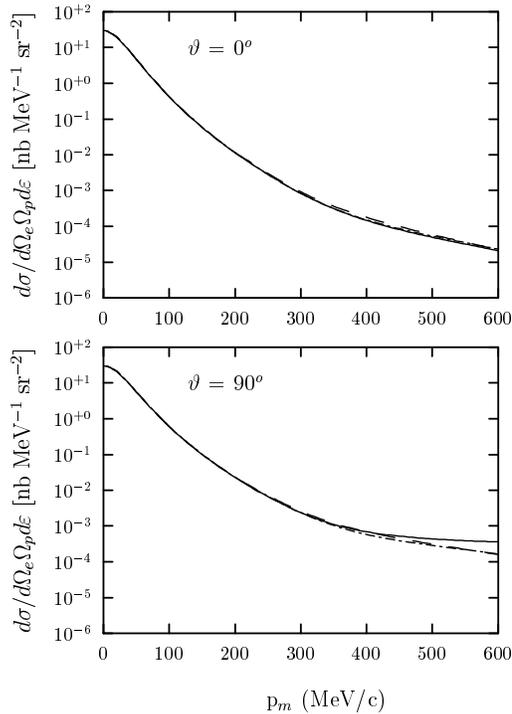, height=20cm}
\vspace{-8cm}
\end{center}
\caption{The differential cross section for the $^2H(e,e'p)n$ reaction
is shown for fixed kinetic energy, $T_{kin} = 1$ GeV, of the outgoing
proton in parallel and perpendicular kinematics. The curves have the same 
meaning as in Fig. \ref{fig2}.}
\label{fig2a}
\end{figure}

In order to be as precise as possible in our investigation of
relativistic effects, we compare the fully relativistic calculation
(Model R), the Argonne V18 plus full relativistic current (Model A1),
and the Gross S wave and D wave plus full relativistic current (Model
G1). The curves are shown in Figs. \ref{fig2} and \ref{fig2a}. The
kinematic conditions are the same as in
Figs. \ref{fig1} and \ref{fig1a}. First, we discuss the results for fixed
energy and momentum transfer. Due to the fact that the momentum
distribution calculated with the Gross wave function is slightly lower
than the Argonne V18 momentum distribution, the Gross S+D plus full
relativistic current results are slightly lower than the AV18 plus
full relativistic current results. This improves the agreement with
the full relativistic calculation for the lowest energy transfer,
$\omega = 0.61$ GeV for momenta less than 400 MeV/c. For higher
missing momenta, the deviation from the full relativistic calculations
increases a bit, although not significantly, leaving a little more
room for genuine relativistic nuclear dynamics effects.  In
Fig. \ref{fig2a}, we show the results for fixed kinetic energy of the
outgoing proton. For parallel kinematics, the use of the Gross S and D
waves plus full relativistic current results improves the agreement
with the fully relativistic calculation, the two curves differ only
slightly.  For perpendicular kinematics, all three curves agree nicely
for missing momenta up to 350 MeV/c, then the Gross S and D wave plus
full relativistic current calculation differs from the full
calculation a bit more than the corresponding AV18 curve.

So, overall, we have very good agreement between the fully
relativistic calculation and the AV18 or Gross S and D wave plus
full relativistic current. In contrast to this, a calculation with
the strict non-relativistic reduction of the electromagnetic current
disagrees markedly with the relativistic results.
In the remainder of this paper, we discuss the possible sources for the
remaining small discrepancies at higher missing momenta.

\subsection{P-wave contributions}

\begin{figure}
\begin{center}
\leavevmode
\epsfig{file = 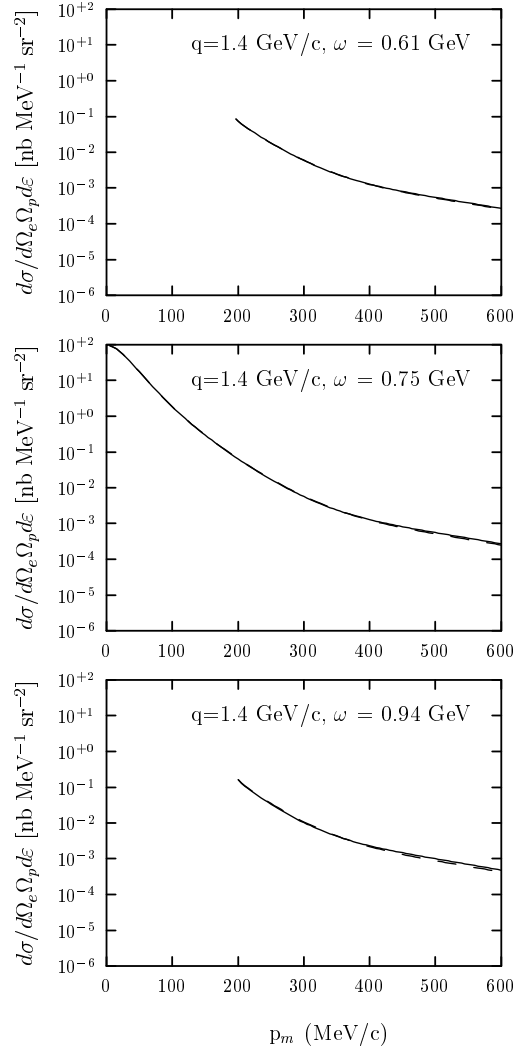, height=20cm}
\end{center}
\vspace{-3cm}
\caption{The differential cross section for the $^2H(e,e'p)n$ reaction
is shown for fixed 3-momentum transfer $q = 1.4$ GeV/c and different
values of the fixed energy transfers $\omega$: 0.61 GeV (a), 0.75 GeV
(b), and 0.94 GeV (c).  The kinematic conditions in the middle panels
correspond to quasi-free conditions. The solid line shows the result of
the fully relativistic calculation (Model R), the dashed
line shows the result of the fully relativistic calculation without 
the P waves (Model RSD).}
\label{fig3}
\end{figure}

\begin{figure}
\begin{center}
\vspace{-3cm}
\leavevmode
\epsfig{file = 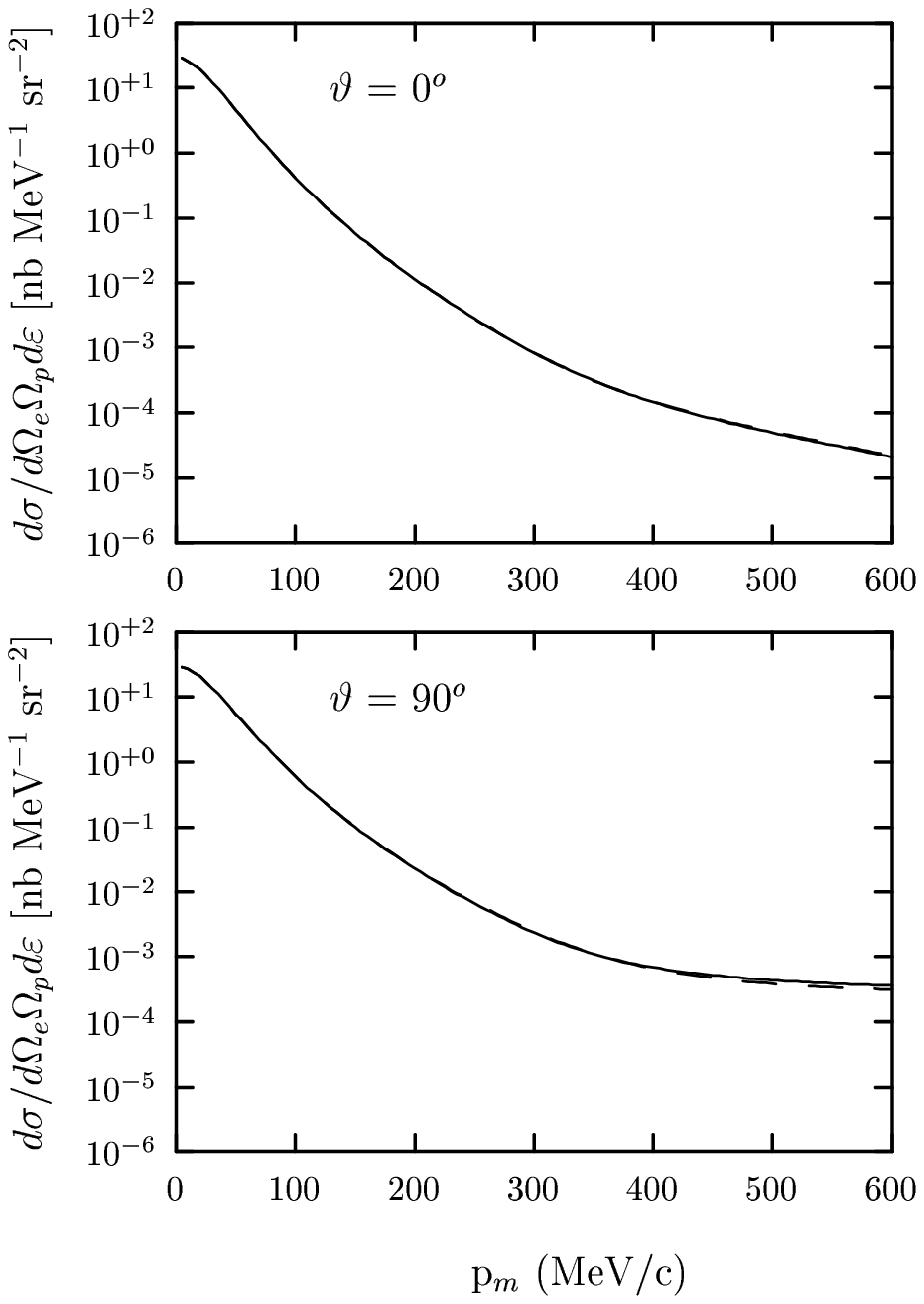, height=20cm}
\vspace{-9cm}
\end{center}
\caption{The differential cross section for the $^2H(e,e'p)n$ reaction
is shown for fixed kinetic energy, $T_{kin} = 1$ GeV, of the outgoing
proton in parallel and perpendicular kinematics. The curves have the same 
meaning as in Fig. \ref{fig3}.}
\label{fig3a}
\end{figure}

 One genuinely relativistic effect that cannot be taken care of by the
electromagnetic current is the emergence of P waves in the deuteron
wave function. They arise naturally from the lower spinor components,
and there is a P-wave singlet and a P-wave triplet contribution, see
e.g. \cite{buckgross}.  Although their normalization is extremely
small compared to the S-wave and the D-wave, they can play an
important role, e.g.  in the calculation of the elastic deuteron
form-factors, where they shift the location of the minimum of $B(Q^2)$
by a considerable amount \cite{eldeuff}. How important are they for
$(e,e'p)$ reactions?  We investigate this question by switching off
the P-wave contribution and comparing the full result (Model R) with
the result without P-waves (Model RSD) in
Figs. \ref{fig3} and \ref{fig3a}.  Overall, the differences between the
full calculation and the calculation without P waves are very small.
In the fixed q and $\omega$ kinematics, the two curves coincide for
the lowest energy transfer of 0.61 GeV, and barely differ for $\omega
= 0.75$ GeV. Only for the highest energy transfer of 0.94 GeV, the P
waves have an effect, their inclusion slightly increases the
result. We would like to point out that this is due to interference
between the P waves and the S and D waves.  The contribution of the P
waves alone is approximately two orders of magnitude smaller than the
contribution of S and D waves, so the interference effects must be
quite large in order to be visible.  In Fig. \ref{fig3a}, in parallel
kinematics (top panel), the two curves barely differ. It is
interesting, though, to see that the interference in this case is
destructive, leading to a slightly enhanced result when the P waves
are omitted. In perpendicular kinematics, the omission of the P waves
lowers the result slightly for missing momenta higher than 400
MeV/c. In each case, when comparing with Figs. \ref{fig1} and  \ref{fig1a}
and with Figs. \ref{fig2} and \ref{fig2a}, one sees that the P wave
contribution explains a part of the difference between the fully
relativistic result and the other calculations, but not all of it.
Another important aspect for calculations at higher missing momenta
are off-shell prescriptions, they will be discussed in the next
section.

One additional remark is in order: although the P-waves are only of
marginal importance for the calculation of the differential cross
section and the large responses, they may play a more important role
in the calculation of single or double polarization observables, by
interfering with one of the larger contributions. Similar situations
arise e.g. when considering the second-order convective spin-orbit
contribution to the transverse-transverse response function, which is
considerable due to the interference with the magnetization current,
the largest current component that cannot contribute to the
transverse-transverse response function otherwise \cite{relcur}; or
when considering the contribution of the spin-orbit final state
interaction to the fifth response function, which allows for a large
interference of the two dominant components of the current that is not
present otherwise \cite{sofsi}. In general, one needs to be careful
in situations where a change of the spin structure will allow for
a new, large contribution. In these situations, it is advisable to take
a close look at the relevant Clebsch-Gordan coefficients and to determine
if additional important contributions could arise from the presence of
P-waves.

\subsection{Off-shell effects}

\begin{figure}
\begin{center}
\leavevmode
\epsfig{file = 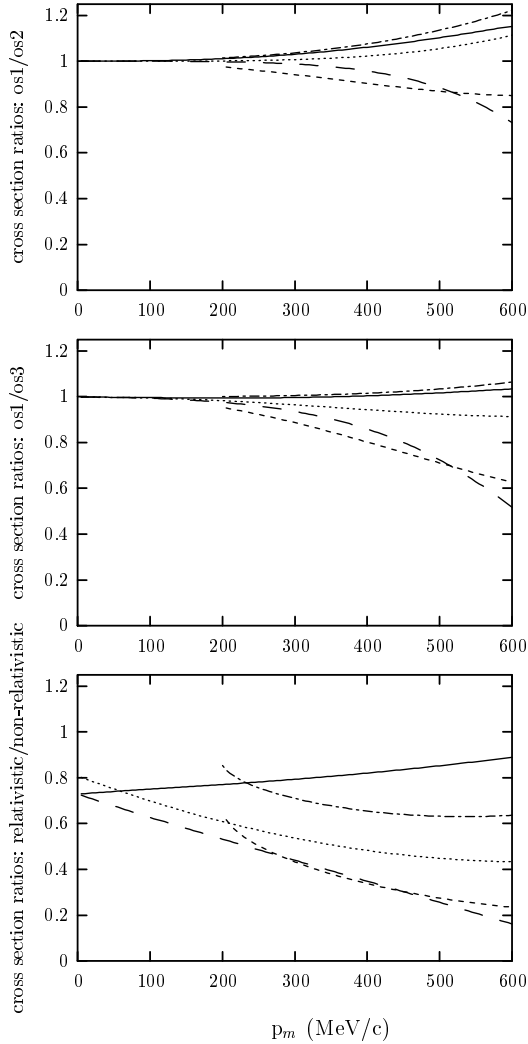, height=20cm}
\end{center}
\vspace{-3cm}
\caption{The upper panel shows the ratio of the off-shell electron
proton cross sections calculated with prescription 1 and prescription
2.  The different lines correspond to different kinematic settings:
the solid line shows the ratio for $T_{kin} = 1$ GeV, parallel
kinematics, the long-dashed line for $T_{kin} = 1$ GeV, perpendicular
kinematics, the dash-dotted line for $q = 1.4$ GeV/c, $\omega = 0.61$
GeV, the dotted line for $q = 1.4$ GeV/c, $\omega = 0.75$ GeV, and the
short-dashed line for $q = 1.4$ GeV/c, $\omega = 0.94$ GeV.  In the
middle and the lower panel, the different lines correspond to the same
kinematic conditions as in the upper panel of this figure.  The middle
panel shows the ratio of the off-shell electron proton cross sections
calculated with prescription 1 and prescription 3.  The lower panel
shows the ratio of the off-shell electron proton cross sections
calculated with the full relativistic current and the strict
non-relativistic reduction of the current for a fixed off-shell
prescription, namely off-shell 1. }
\label{figwqos}
\end{figure}

Whenever one applies the one-body electromagnetic current operator not
to a free nucleon, but to a nucleon bound in the nucleus, one needs to
introduce an off-shell prescription.  Nucleons in the nucleus are
bound and therefore off-shell; they do not fulfill the same
energy-momentum relations as free nucleons.  Currently, there exists
no microscopic description of this off-shell behavior that can be
applied for a wide range of kinematic conditions --- there are only ad
hoc prescriptions, which lead to differing results for certain
kinematics \cite{deforest,koch,cdp}. The variations tend to increase
with increasing momentum of the initial nucleon - the higher its
momentum, the further it is off-shell.  The fully relativistic
calculation (Model R, Model RSD) uses a more sophisticated off-shell
prescription, which takes into account the actual kinematics.  In the
calculation with the non-relativistic wave functions and the exact
form of the current operator, we have chosen the popular ansatz of
employing the ``on-shell form'' of the current. However, these are
specific choices among an infinite number of prescriptions.  We will
quantify the theoretical uncertainty due to these choices in the
following. It is easy to see why there are, in principle, infinitely
many possibilities to choose an off-shell prescription. Consider the
case of non-relativistic PWIA, and of factorization holding there,
i.e. we are investigating the off-shell effects in the framework of
non-relativistic wave function and full current operator now. Due to
the factorization, the cross section can be described by the product
of momentum distribution and off-shell electron-proton cross section,
see eq. (\ref{pwiafactor}). For a free nucleon, one can write down the
$ep$ cross section, and then rewrite it using simple algebra, that in
general will employ the on-shell energy momentum relation for the free
nucleon. On-shell, all these ways of writing the $ep$ cross section
are equivalent - off-shell, they differ, and of course there is no way
to identify a single expression as the correct one. E.g., one obtains
different versions of the off-shell $ep$ cross section if one starts
out from the one-body current in the form given in eq. (\ref{defcur}),
or if one uses the Gordon decomposition on the current. This is
precisely how deForest \cite{deforest} obtained his off-shell
prescription cc1 (current like in eq. (\ref{defcur})) and cc2 (Gordon
decomposition applied to the current).  At first sight, the situation
might look rather grim, however, things are not as bad as one might
assume. In the following, we compare the employed off-shell
prescriptions. In eq. (\ref{defresp}), we gave the expressions for the
single nucleon responses in terms of the $\xi$s, as they have been
used in the calculations shown above (Models A1 and G1), and also in
terms of $W_1, W_2$ (Model A2).  This latter form was obtained by
doing some algebra involving the on-shell kinematic relations, and
therefore constitutes a different off-shell prescription. In
Fig. \ref{figwqos}, top panel, we show the ratio of the $ep$ cross
section calculated with the $\xi$s, called off-shell prescription 1,
and with $W_1, W_2$, referred to as off-shell prescription 2. The
curves show the ratios for the kinematic settings discussed in this
paper.  One sees that, according to expectations, the off-shell
prescriptions agree for momenta below 200 MeV/c, and then start to
diverge. At the highest missing momenta considered here, the
deviations range from 10\% to 25\%. At the intermediate momenta, the
typical deviation from 1 is about 5 \%.  The largest deviations occur
for $T_{kin} = 1 GeV$ in perpendicular kinematics (long dashed line),
and for the $q = 1.4$ GeV/c, $\omega = 0.61$ GeV case (dash-dotted
line).  In the middle panel of Fig. \ref{figwqos}, we show the ratio
of off-shell prescription 1 and the off-shell prescription used in the
covariant calculation, denoted as off-shell 3. The explicit formulas
for off-shell 3 are given in appendix \ref{apb}. Again, the two
off-shell prescriptions agree nicely for low missing momenta, and
start to diverge for $p_m > 200$ MeV/c. The deviations are large,
almost 50 \% at the highest missing momenta for $T_{kin} = 1 GeV$ in
perpendicular kinematics, and also considerable at high missing
momenta for $q = 1.4$ GeV/c, $\omega = 0.61$ GeV. In all other
considered kinematics, the deviations do not exceed 5\% even at the
highest missing momenta.  Note that the deviations from $1$ for the
ratio of off-shell 2 to off-shell 3 are smaller than any of the ratios
shown here, we have chosen to display the ``extreme'' cases here.

Although the off-shell prescriptions 1,2, and 3 are all relativistic,
there remains an ambiguity in the choice of the off-shell
kinematics. The scope of this paper was to investigate the treatment
of relativistic effects, not the specific off-shell choices.  However,
our results give a good idea about the size of the uncertainties, and
about the kinematic regions which are sensitive to them; whereas in
parallel kinematics, the results are quite insensitive to the employed
off-shell prescription, the results in perpendicular kinematics are
more sensitive. Below missing momenta of $200$ MeV/c, off-shell 
ambiguities practically do not exist, and for $p_m < 400$ MeV/c,
they will in general not be larger than 10 \%.

Sometimes, the full relativistic current is rejected as introducing
off-shell contributions, and the claim is made that the
non-relativistic reduction is preferable as it would avoid such
ambiguities.  This argument is not correct. Firstly, even the
non-relativistic version of the current contains a dependence on the
nucleon momentum, leading to terms proportional to the perpendicular
momentum squared in the single nucleon responses, see
eq. (\ref{snrnr}).  On-shell, this momentum can be replaced with the
energy by using the on-shell kinematic relations, and therefore one
does have off-shell ambiguities even in this case. One might argue
that the non-relativistic versions of the charge operator and the
magnetization current do not contain any dependence on the initial
nucleon momentum, and should therefore be free of off-shell
ambiguities. Here, one needs to ask, however, if this feature was not
bought at a too high price: the $ep$ cross sections calculated in the
strict non-relativistic reduction differ from the relativistic version
by 30 \% to 80 \%, and this discrepancy starts at $p_m = 0$ MeV/c. We
demonstrate this point by presenting the ratio of the relativistic to
the non-relativistic $e-p$ cross section for one specific off-shell
prescription, namely the on-shell form with the $\xi$s, referred to as
off-shell 1, see Fig. \ref{figwqos}, lower panel. 
Even though the off-shell effects in the full relativistic version of
the current do introduce a theoretical uncertainty at higher missing
momenta, the relativistic treatment is definitely superior to the
non-relativistic version, where one knows that the results are off for
all missing momentum values, and they are off up to a factor of 5.

\section{Summary}

The principal goal of this paper has been to investigate the origin of
the relativistic effects in the electrodisintegration of the deuteron.
Apart from the obviously relativistic kinematics, relativistic effects
occur in the electromagnetic current operator and the nuclear
dynamics. We have shown that calculations with a non-relativistic wave
function and the fully relativistic current operator reproduce the
fully relativistic calculation very well up to $400$ MeV/c of missing
momentum in all kinematics. At the same time, a calculation with a
non-relativistic wave function and a strictly non-relativistic current
operator drastically fails to reproduce the relativistic results, in
all kinematics and even for the lowest missing momenta.

We have shown that the relativistic effects in the nuclear dynamics,
namely, the presence of P-waves, are rather small in the cross
section. They do appear only in places where one would expect
relativistic effects to show up, i.e. for high missing momenta, $p_m >
400$ MeV/c and for energy transfers which are large and comparable to
the momentum transfer.  The remaining discrepancy that we found in a
few kinematics for high missing momenta was shown to stem from the
different off-shell prescriptions used in the two calculations. 

Since
\cite{pvow} shows that the effects of bound state dynamical relativity
are small also in heavier nuclei, the technique of \cite{relcur} can
reliably be applied to heavier nuclei, too.

These findings verify the conjecture of \cite{relcur} that at energies
of a few GeV, the bulk of the relativistic effects in electronuclear
reactions stems from the current operator, and not from the nuclear
dynamics.

\acknowledgments 
The authors thank J. Adam, T. W. Donnelly, and
F. Gross for many stimulating discussions on the subject of
relativity.  The authors thank R. Schiavilla for providing a
parameterization of the Argonne V18 and CD Bonn deuteron wave
functions.  This work was in part supported by funds provided by the
U.S. Department of Energy (D.O.E.)  under cooperative research
agreement \#DE-AC05-84ER40150.

\appendix
\section{Overview over presented calculations: notation}
\label{apnames}

For the convenience of the reader, we list all calculations and
assign a short name for them:

\begin{tabular}{cccc}
wave function & responses/off-shell prescription & description & name
\\ \\ [0.5ex] \hline \hline 
Gross wave function; & as in appendix \ref{apa}, & fully relativistic,&
Model R \\ 
S, D, P waves & off-shell 3 & covariant
calculation &\\ 
\hline 
Gross wave function; & as in appendix \ref{apa}, & fully relativistic,&
Model RSD \\ 
S, D waves & off-shell 3 & covariant
calculation &\\ 
\hline 
Argonne V18; & as in eq. (\ref{snrex}), & AV18 plus full&
Model A1 \\
 S, D waves & with the $\xi$,  off-shell 1 & relativistic current &
\\
\hline 
Argonne V18; & as in eq. (\ref{snrex}), & AV18 plus full &
 Model A2 \\ S, D waves &  with $W_1, W_2$, off-shell 2 &
 relativistic current & \\
\hline 
Gross wave function; & as in eq. (\ref{snrex}), & Gross S and D
wave function & Model G1 \\ S, D waves & with the $\xi$, off-shell 1 &
and full relativistic current &\\
\hline 
Argonne V18; & as in eq. (\ref{snrnr}), & AV18 plus strictly &
Model NR \\S, D waves & with the $\xi$, off-shell 1 & non-relativistic
current & \\
\hline 
\end{tabular}

\section{Fully relativistic nuclear responses}
\label{apa}

Here, the explicit analytic expressions for the nuclear response
functions using the off-shell prescription of the fully relativistic,
covariant calculation are given:

\begin{eqnarray}
R_L &=& \frac{1}{8 \pi} \, \frac{q ^2}{Q ^2} \, \{ (C_1^{TV}
 n_{TV} (p) + C_1^{SV} n_{SV} (p) + C_1^{S} n_{S} (p)) \nonumber \\ 
&&
 + C_2^{SV} n_{SV} (p) \, [ (M_d - E_p) \frac{q}{Q} - \frac{\omega}{Q}
 p_z]^2 \nonumber \\ 
&&
 + (C_3^{TV} n_{TV} (p) + C_3^{SV} n_{SV} (p)) \,
 2 M_d \, \frac{q}{Q} \, [(M_d - E_p) \frac{q}{Q} - \frac{\omega}{Q} p_z] \}
\\
%\end{eqnarray}
%
%\begin{eqnarray}
R_T  &=& \frac{1}{8 \pi} \, \, \{ - 2 \, (C_1^{TV}
 n_{TV} (p) + C_1^{SV} n_{SV} (p) + C_1^{S} n_{S} (p)) \nonumber \\ 
&&
 + C_2^{SV} n_{SV} (p) \, \,  p_{\perp}^2 \}\\
%\end{eqnarray}
%
%
%\begin{eqnarray}
R_{TT} &=& -\frac{1}{8 \pi} \,  \, C_2^{SV} n_{SV} (p) \, \, p_{\perp}^2
 \, \cos (2 \varphi)\\
%\end{eqnarray}
%
%\begin{eqnarray}
R_{TL} &=& \frac{1}{8 \pi} \, \frac{q}{Q} \, \{C_2^{SV} n_{SV} (p)
\, 2 \sqrt{2} \, [ (M_d - E_p) \frac{q}{Q} - \frac{\omega}{Q}
 p_z]\,  p_{\perp} \cos (\varphi) \nonumber \\
&&
+ (C_3^{TV} n_{TV} (p) + C_3^{SV} n_{SV} (p)) \, 2 \sqrt{2} \, 
M_d \, \frac{q}{Q}  \, p_{\perp} \cos (\varphi) \}
\end{eqnarray}

The coefficients $C$ of the time-like vector, space-like vector, and
scalar momentum distributions are given by:

\begin{eqnarray}
C_1^{TV} &=& \frac{\Delta  - M_d^2}{4 \, m^2 \, M_d^2} \, \{(M_d^2 + \Delta)
( F_1^2(Q^2) + \tau F_2^2 (Q^2)) \nonumber \\
& & + 2 M_d \, \omega \, F_1 (Q^2) \, (F_1 (Q^2)
+ F_2 (Q^2)) - \frac{M_d \, \Delta \, \omega}{2 \, m^2} F_2^2 (Q^2) \} \\
%\end{eqnarray}
%
%\begin{eqnarray}
C_2^{TV} &=& 0 \\
%\end{equation}
%
%\begin{equation}
C_3^{TV} &=& \frac{M_d^2 - \Delta}{2 \, m^2 \, M_d^2} 
( F_1^2(Q^2) + \tau F_2^2 (Q^2)) \\
%\end{equation}
%
%\begin{equation}
C_4^{TV} &=& 0 \,
\end{eqnarray}
where $\Delta = p_1^2 - m^2 = M_d^2 - 2 M_d E_p$ measures the
``off-shellness'' of the bound nucleon.

\begin{eqnarray}
C_1^{SV} &=& - \left ( 1 - \frac{M_d^2}{4 \, m^2} \right )
( F_1^2(Q^2) + \tau F_2^2 (Q^2)) - 2 \tau  F_1 (Q^2) (F_1 (Q^2)
+ F_2 (Q^2)) \nonumber \\
& &+ \frac{\Delta}{2\,  m^2} F_1 (Q^2) F_2 (Q^2) - \frac{\Delta^2}
{8\,  m^4}  F_2^2 (Q^2) + \frac{\Delta^2}{4 \, m^2 \, M_d^2} 
( F_1^2(Q^2) + \tau F_2^2 (Q^2)) \nonumber \\
&& - \frac{\Delta^2 \omega}{8 \, m^4 \, M_d}
 F_2^2 (Q^2) + \frac{\Delta \omega}{2 \, m^2 \, M_d} ( F_1^2 (Q^2)
+ 2  F_1 (Q^2)  F_2 (Q^2) - \frac{M_d^2}{4 \, m^2}  F_2^2 (Q^2))
\nonumber \\
& & + \frac{M_d \, \omega}{2 \, m^2} F_1 (Q^2) (F_1 (Q^2)
+ F_2 (Q^2)) \\
%\end{eqnarray}
%
%
%\begin{equation}
C_2^{SV} &=& \frac{2}{m^2} \, ( F_1^2(Q^2) + \tau F_2^2 (Q^2)) \\
%\end{equation}
%
%\begin{equation}
C_3^{SV} &=& - \frac{\Delta + M_d^2}{2 \, m^2 \, M_d^2}
 (F_1^2(Q^2) + \tau F_2^2 (Q^2)) \\
%\end{equation}
%
%\begin{equation}
C_4^{SV} &=& 0 \\
%\end{equation}
%
%\begin{equation}
C_1^{S} &=&  F_1^2(Q^2) - \tau F_2^2 (Q^2) - 2 \tau F_1(Q^2) F_2(Q^2)
 - \frac{\Delta}{2 \, m^2}  F_1(Q^2) F_2(Q^2) \\
%\end{equation}
%
%\begin{equation}
C_2^{S} &=& 0 \\
%\end{equation}
%
%
%\begin{equation}
C_3^{S} &=& 0 \\
%\end{equation}
%
%\begin{equation}
C_4^{S} &=& 0 
\end{eqnarray}

Note that the responses given here are to be inserted into
eq. (\ref{wqdef}) for the cross section in the lab frame. The responses
$\tilde R$ used in \cite{dmtrgross} in eq. (95) are connected to these
responses in the following way:

\begin{eqnarray}
R_L &=& \frac{2 \pi ^2}{m^2} \frac{W}{M_T} \frac{q^2}{Q^2} \tilde R_L
\nonumber \\
R_T &=& \frac{2 \pi ^2}{m^2} \frac{W}{M_T}  \tilde R_T \nonumber \\
R_{TT} &=& \frac{2 \pi ^2}{m^2} \frac{W}{M_T}  \tilde R_{TT} 
\cos (2 \varphi)
\nonumber \\
R_{TL} &=& \frac{2 \pi ^2}{m^2} \frac{W}{M_T} \frac{q}{Q} \tilde R_{TL}
\cos (\varphi)
\end{eqnarray}

Here, $W$ indicates the invariant mass of the final state, and $M_T$ is
the target mass.

\section{Single nucleon off-shell responses}
\label{apb}

In the case of vanishing P-wave contributions, the different momentum
distributions coincide:
\begin{equation}
n_{TV} (p) = n_{SV} (p) = n_{S} (p) \equiv n^+ (p) = u^2 (p) + w^2 (p)
\end{equation}
Note that the normalization condition of the Gross wave function as
given in eq. (\ref{wallynormcond}) differs from the one employed for
the Bonn and Paris wave function, eq. (\ref{nrnorm}). The different
normalizations have of course been taken into account when we compared
the two results.  For vanishing P-waves, the responses factorize, and
the resulting off-shell single nucleon responses are called
``off-shell 3'' in the text. For the convenience of the reader, they
are given here explicitly:

\begin{eqnarray}
R_L^{sn, no \, p} &=& \frac{1}{8 \pi} \, \frac{q ^2}{Q ^2} \, \{ -2 \tau
(F_1 (Q^2) + F_2 (Q^2))^2 + \frac{\Delta^2}{2 \, m^2 \, M_d^2} (
F_1^2(Q^2) + \tau F_2^2 (Q^2)) - \frac{\Delta^2}{8 \, m^4} F_2^2 (Q^2)
\nonumber \\ & & - \frac{\Delta^2 \, \omega}{4 \, m^4 \, M_d^2} \, F_2^2
(Q^2) + \frac{\Delta \, \omega}{ m^2 \, M_d} F_1 (Q^2) (F_1 (Q^2) +
F_2 (Q^2)) \nonumber \\ & &
+ \frac{2}{m^2} (F_1 (Q^2) + F_2 (Q^2))^2 [ (M_d - E_p)
\frac{q}{Q} - \frac{\omega}{Q} p_z]^2 \nonumber \\ & &
-\frac{\Delta}{m^2 \, M_d^2} ( F_1^2(Q^2) + \tau F_2^2 (Q^2)) \, 2 M_d
\, \frac{q}{Q} \, [(M_d - E_p) \frac{q}{Q} - \frac{\omega}{Q} p_z] \} \\
%\end{eqnarray}
%
%\begin{eqnarray}
R_T^{sn, no \, p} &=& \frac{1}{8 \pi} \, \{ 4 \tau (F_1 (Q^2) + F_2
(Q^2))^2 - \frac{\Delta}{m^2 \, M_d^2} (F_1^2(Q^2) + \tau F_2^2 (Q^2))
\nonumber \\ & & + \frac{\Delta^2}{4 \, m^4} F_2^2 (Q^2) +
\frac{\Delta^2 \, \omega} {2 \, m^4 M_d} F_2^2 (Q^2) - \frac{ 2 \Delta
\omega}{m^2 \, M_d^2} F_1 (Q^2) (F_1 (Q^2) + F_2 (Q^2)) 
\nonumber \\ & & + \frac{2}{m^2}
(F_1^2(Q^2) + \tau F_2^2 (Q^2)) p_{\perp}^2 \} \\
%\end{eqnarray}
%
%\begin{equation}
R_{TT}^{sn, no \, p} &=& - \frac{1}{8 \pi} \,  \frac{2}{m^2}
(F_1^2(Q^2) + \tau F_2^2 (Q^2)) p_{\perp}^2 \, \cos (2 \varphi) \\
%\end{equation}
%
%\begin{eqnarray}
R_{TL}^{sn, no \, p} &=& \frac{1}{8 \pi} \, \frac{q}{Q} \, \{
\frac{2}{m^2} (F_1^2(Q^2) + \tau F_2^2 (Q^2)) \, 2 \sqrt{2} \, [ (M_d
- E_p) \frac{q}{Q} - \frac{\omega}{Q} p_z]\, p_{\perp}
\nonumber \\ & & - 
\frac{\Delta}{m^2 \, M_d^2} (F_1^2(Q^2) + \tau F_2^2 (Q^2))
\, 2 \sqrt{2} \, M_d \, \frac{q}{Q} p_{\perp} \} \cos(\varphi)
\end{eqnarray}

% = = = = = = = = = = = = = = = = = = = = = = = = = = = = = = = = = = =


\begin{thebibliography}{299}

\bibitem{relcur} 
S. Jeschonnek and T.W. Donnelly, {\sl Phys. Rev. C} {\bf
57}, 2438  (1998).

\bibitem{vano}
J. W. Van Orden, N. Devine, and F. Gross, {\sl Phys. Rev. Lett.}
{\bf 75},  4369 (1995). 

\bibitem{tjon}
E. Hummel and J. A. Tjon, {\sl Phys. Rev. C} {\bf 49},  21 (1994).
\bibitem{pvquique}
J. E. Amaro, J. A. Caballero, T. W. Donnelly, A. M. Lallena,
E. Moya de Guerra, and J. M. Ud\'{\i}as, {\sl Nucl. Phys.} {\bf A602},
 263 (1996). 

\bibitem{quiqueinc}
J. E. Amaro, J. A. Caballero, T. W. Donnelly, and E. Moya de Guerra,
{\sl Nucl. Phys.} {\bf A611},  163 (1996).

\bibitem{quiquecoin} 
J. E. Amaro and T. W. Donnelly, 
{\sl Ann. Phys.} {\bf 263},  56 (1998).


\bibitem{sofsi}
S. Jeschonnek and T.W. Donnelly, {\sl Phys. Rev. C} {\bf 59},
 2676 (1999).

\bibitem{raskintwd}
A. S. Raskin and T. W. Donnelly, {\sl Ann. of Phys.} {\bf 191},
78 (1989).

\bibitem{grosseq}
F. Gross, {\sl Phys. Rev.} {\bf 186}, 1448 (1969);
F. Gross, {\sl Phys. Rev. D} {\bf 10}, 223 (1974);
F. Gross, {\sl Phys. Rev. C} {\bf 26}, 2203 (1982).

\bibitem{dmtrgross}
V. Dmitrasinovic and F. Gross, {\sl Phys. Rev. C} {\bf40},
2479 (1989).

\bibitem{wallynf}
A. Picklesimer and J. W. Van Orden, {\sl Phys. Rev. C} {\bf40},
290 (1989).

\bibitem{rpwia} 
J. A Caballero, T. W. Donnelly, E. Moya de Guerra,
J. M. Ud\'{\i}as, {\sl Nucl. Phys.} {\bf A632}, 323 (1998).

\bibitem{gardner}
S. Gardner and J. Piekarewicz, 
{\sl Phys. Rev. C} {\bf 50}, 2822  (1994).

\bibitem{av18}
R. B. Wiringa, V. G. J. Stoks, and R. Schiavilla,
{\sl Phys. Rev. C} {\bf 51}, 38  (1995).

\bibitem{yscal}
D. B. Day, J. S. McCarthy, T. W. Donnelly, and I. Sick,
{\sl Annu. Rev. Nucl. Part. Sci.} {\bf 40},  357 (1990).

\bibitem{bonn}
R. Machleidt, K. Holinde and C. Elster,
{\em Phys. Rep.} {\bf 149}, 1 (1987).

\bibitem{paris}
M. Lacombe, B. Loiseau, R. Mau, J. C\^ot\`e, P. Pir\`es and R. Tourreil,
{\em Phys. Lett.} {\bf B101}, 139 (1981). 


\bibitem{machleidt}
R. Machleidt, F. Sammarruca, and Y. Song,
{\sl Phys. Rev. C} {\bf 53}, 1483 (1996).


\bibitem{buckgross}
W. W. Buck and F. Gross, {\sl Phys. Rev. D} {\bf 20},
 2361 (1979).

\bibitem{eldeuff}
J. W. Van Orden, N. Devine, and F. Gross, {\sl Phys. Rev. Lett.}
{\bf 75} 4369 (1995).

\bibitem{deforest} 
T. de Forest Jr.,
{\sl Nucl.Phys.} {\bf A392},  232 (1983).
 
\bibitem{koch}
H. W. L. Naus, S. J. Pollock, J. H. Koch, and U. Oelfke,
{\sl Nucl. Phys.} {\bf A509}, 717  (1990).

\bibitem{cdp}
J. A. Caballero, T. W. Donnelly, and G. I. Poulis,
{\sl Nucl. Phys.} {\bf A555}, 709  (1993).

\bibitem{pvow}
A. Picklesimer, J. W. Van Orden, and S. J. Wallace,
{\sl Phys. Rev. C} {\bf 32} 1312 (1985).


\end{thebibliography}
\end{document}